\documentclass[conference,a4paper]{IEEEtran}
\IEEEoverridecommandlockouts

\usepackage{cite}
\usepackage{dblfloatfix}
\usepackage{subcaption}
\usepackage{overpic}
\usepackage{amsmath,amssymb,amsfonts}
\usepackage{graphicx}
\usepackage{textcomp}
\usepackage{xcolor}
\usepackage{float}
\usepackage{amsthm}
\usepackage{graphicx}
\usepackage{epstopdf}
\usepackage{amsmath,bm}
\usepackage{amsfonts}
\usepackage{amssymb}
\usepackage{color}
\usepackage{multirow}
\usepackage{multicol}
\usepackage{soul,xcolor}
\usepackage{algorithm}
\usepackage{algpseudocode}

\usepackage{comment}

\theoremstyle{plain}

\newcommand{\vect}[1]{\mathbf{#1}}

\def\Htran{\mbox{\tiny $\mathrm{H}$}}
\def\Ttran{\mbox{\tiny $\mathrm{T}$}}
\def\CN{\mathcal{N}_{\mathbb{C}}} 
\def\imaginary{\mathsf{j}} 
\def\imagunit{\imaginary}

\def\H{{\bf H}}

\def\N0{N_0}
\def\He2e{\H_{\textrm{e2e}}}

\IEEEoverridecommandlockouts

\begin{document}

\title{Impact of Phase Noise and Power Amplifier Non-Linearities on Downlink Cell-Free Massive MIMO-OFDM Systems }

\author{\IEEEauthorblockN{
\"Ozlem Tu\u{g}fe Demir\IEEEauthorrefmark{1} and   
Emil Bj{\"o}rnson\IEEEauthorrefmark{2}  
}                                     
\IEEEauthorblockA{\IEEEauthorrefmark{1}
Department of Electrical-Electronics Engineering, TOBB ETÜ, Ankara, T\"urkiye, ozlemtugfedemir@etu.edu.tr}
\IEEEauthorblockA{\IEEEauthorrefmark{2}
Department of Computer Science, KTH Royal Institute of Technology, Stockholm, Sweden, emilbjo@kth.se}
\thanks{\"O. T. Demir was supported by 2232-B International Fellowship for Early Stage Researchers Programme funded by the Scientific and Technological Research Council of T\"urkiye. E.~Bj\"ornson was supported by the SUCCESS project funded by the Swedish Foundation for Strategic Research.}
}

\maketitle
\begin{abstract}

Cell-free massive MIMO (multiple-input multiple-output) is a key enabler for the sixth generation (6G) of mobile networks, offering significant spectral and energy efficiency gains through user-centric operation of distributed access points (APs). However, its reliance on low-cost APs introduces inevitable hardware impairments, whose combined impact on wideband downlink systems remains unexplored when analyzed using behavioral models. This paper presents a comprehensive analysis of the downlink spectral efficiency (SE) in cell-free massive MIMO-OFDM systems under practical hardware impairments, including phase noise and third-order power amplifier nonlinearities. Both centralized and distributed precoding strategies are examined. By leveraging the Bussgang decomposition, we derive an SE expression and quantify the relative impact of impairments through simulations. Our results reveal that phase noise causes more severe degradation than power amplifier distortions—especially in distributed operation—highlighting the need for future distortion-aware precoding designs.
\end{abstract}


\section{Introduction}
Cell-free massive multiple-input multiple-output (MIMO) has emerged as a foundational technology for sixth-generation (6G) wireless networks, offering significant improvements in spectral and energy efficiency through cooperative, user-centric communication that eliminates traditional cell boundaries~\cite{cell-free-book}. In this architecture, a large number of distributed access points (APs), each equipped with a few antennas, simultaneously serve all user equipments (UEs) over the same time-frequency resources. This distributed topology enhances spatial diversity and mitigates inter-cell interference, delivering nearly uniform quality-of-service across the coverage area.

To achieve scalability and cost efficiency, APs are typically implemented using low-cost hardware components. However, this introduces various radio-frequency (RF) impairments, including phase noise, in-phase/quadrature imbalance (IQI), non-linearities in power amplifiers, and limited-resolution analog-to-digital converters (ADCs) and digital-to-analog converters (DACs). Although the impact of these impairments has been extensively analyzed in cellular MIMO systems, their effect on wideband cell-free massive MIMO systems remains insufficiently explored, particularly in the downlink.

Some studies have addressed hardware impairments in the uplink of cell-free systems~\cite{8891922}. However, the majority of existing works adopt narrowband models, which assume frequency-flat channels and fail to capture the complexity of wideband systems that employ orthogonal frequency division multiplexing (OFDM). Notable exceptions include~\cite{10437146}, which investigates the effect of phase noise in OFDM, and~\cite{9576714}, which examines power amplifier non-linearities. Nonetheless,~\cite{10437146} is limited to the uplink scenario. To the best of our knowledge, no prior work has jointly studied the combined impact of phase noise and power amplifier non-linearities in the downlink of cell-free massive MIMO-OFDM systems.

Furthermore, most existing analyses rely on stochastic additive impairment models for mathematical tractability~\cite{10225319,9103262}. While convenient, these models may not adequately capture the behavior of practical transceivers in OFDM-based systems. In contrast, deterministic behavioral models offer a more accurate characterization of nonlinear impairments, using fewer parameters and without reliance on specific hardware implementations~\cite{Schenk2008a}. In this paper, we adopt such behavioral models to represent residual phase noise and third-order power amplifier non-linearities.

This paper addresses the aforementioned gaps by conducting a comprehensive evaluation of the downlink spectral efficiency (SE) in cell-free massive MIMO-OFDM systems under practical hardware impairments. Both centralized precoding based on global channel state information (CSI) and distributed precoding based on local CSI are considered. By leveraging the Bussgang decomposition, we derive an SE expression and perform simulations to assess the performance trade-offs. To the best of our knowledge, this is the first work that jointly analyzes the impact of phase noise and power amplifier non-linearities in the downlink of cell-free massive MIMO systems over OFDM, using behavioral models. While the modeling tools (e.g., Bussgang decomposition) are established, their integration into a practical wideband setting makes the analysis non-trivial. Furthermore, the study reveals new insights into the relative dominance of different impairment sources under centralized and distributed precoding. These findings are not only technically relevant but also highly practical for guiding distortion-aware design in future 6G systems.

\section{System Model}
We consider the downlink transmission of a cell-free massive MIMO system employing OFDM. The network consists of $L$ APs and $K$ UEs arbitrarily distributed across the coverage area. Each UE is equipped with a single antenna, while each AP employs $N$ antennas. The complex baseband-equivalent frequency-selective channel between each AP and UE pair is represented as a finite impulse response (FIR) filter comprising $R$ equally spaced taps.\footnote{In practice, delay spreads between different AP-UE pairs vary, leading to distinct numbers of taps. However, without loss of generality, we assume a uniform maximum number of taps $R$ for all pairs to simplify the analysis \cite{Jin2020}.} The sampling period is defined as $T_s = 1/B$, where $B$ denotes the total bandwidth divided among the $M$ OFDM subcarriers, resulting in a subcarrier spacing of $B/M$.

The $r$-th channel tap between UE $k$ and AP $l$ is represented by the vector ${\bf h}_{kl}[r]\in \mathbb{C}^{N}$, where $r=0,\ldots,R-1$. The number of OFDM subcarriers satisfies $M\geq R$, and each OFDM symbol consists of $M+R-1$ samples, including a cyclic prefix (CP) of length $R-1$. The channel realizations remain constant within each coherence interval containing $n_{\rm{coh}}$ OFDM symbols, and they vary independently across different coherence intervals \cite{Pitarokoilis2016a}. Consequently, the coherence time is given by $T_{\rm{coh}}=n_{\rm{coh}}(M+R-1)T_s$. The channels follow a correlated Rayleigh fading model, i.e., $\vect{h}_{kl}[r]\sim\mathcal{N}_{\mathbb{C}}({\bf 0},{\bf R}_{kl}[r])$, and are assumed to be independent across different delay taps. The covariance matrix ${\bf{R}}_{kl}[r]\in \mathbb{C}^{N\times N}$ captures the spatial correlation among antennas at AP $l$, as well as large-scale propagation characteristics such as path loss and shadowing \cite{Mollen2016a,Ucuncu2020,Jin2020}. The spatial correlation matrices vary with the tap index $r$, so we have independent but unequally distributed fading between taps. These spatial correlation matrices remain fixed during communication. We further assume perfect channel state information (CSI) is available at each AP, allowing us to isolate and focus solely on the adverse impact of hardware impairments on downlink performance.

All APs are connected to a central processing unit (CPU) via fronthaul links. The CPU serves as the network's central node equipped with significant computational resources. To specifically highlight the impact of hardware impairments at the APs, which function as low-complexity radio units, we assume ideal fronthaul connections without latency or errors.

In cell-free networks, two primary approaches exist for downlink operation: centralized and distributed \cite[Sec. 6.1--6.2]{cell-free-book}. In the centralized approach, representing the most advanced implementation, all tasks related to channel estimation and payload data precoding are executed centrally at the CPU, while APs operate merely as remote radio heads forwarding signals generated by the CPU to the UEs. In this scenario, precoding vectors are computed based on comprehensive, aggregated CSI from all APs, enabling efficient interference mitigation. Conversely, the distributed approach involves APs independently estimating local CSI and performing precoding on data encoded centrally at the CPU. Due to precoding relying solely on local CSI, the distributed operation typically exhibits reduced performance compared to the centralized approach.

We will analyze a single arbitrary OFDM symbol designated for downlink data transmission. The frequency-domain data symbol transmitted to UE $k$ on the $m$-th subcarrier is denoted by $\overline{s}_k[m]$. The corresponding frequency-domain channel between AP $l$ and UE $k$ is represented by
\begin{align}
\overline{\vect{h}}_{kl}[m] = \sum_{r=0}^{R-1}\vect{h}_{kl}[r]e^{-\imagunit \frac{2\pi r m}{M}}, \quad m=0,\dots,M-1. 
\end{align}
In the centralized operation, the precoding vectors are designed using the concatenated frequency-domain channel from all APs, defined as
\begin{align}
\overline{\vect{h}}_k[m] = \begin{bmatrix} \overline{\vect{h}}_{k1}[m] \\ \vdots \\ \overline{\vect{h}}_{kL}[m] \end{bmatrix}. \end{align}
A common choice for the centralized precoding vector $\vect{w}_k[m]$ is regularized zero-forcing (RZF) combined with equal power allocation, given by
\begin{align} 
\vect{w}_k[m] = \sqrt{\frac{\rho_{\rm max}}{K}}\frac{\overline{\vect{w}}_k[m]}{\omega_k[m]}, 
\end{align} 
where
\begin{align}
\overline{\vect{w}}_k[m] = \left(\sum_{i=1}^K\overline{\vect{h}}_i[m]\overline{\vect{h}}_i^{\Htran}[m]+\lambda\vect{I}_{LN}\right)^{-1}\overline{\vect{h}}_k[m], \end{align} 
with the regularization parameter $\lambda>0$ and normalization factor 
\begin{align} 
\omega_k[m] = \max_{l\in{1,\ldots,L}} \left\Vert \overline{\vect{w}}_{kl}[m] \right\Vert, 
\end{align}
where the precoding vector used by AP $l$ at subcarrier $m$ is
\begin{align} \overline{\vect{w}}_{kl}[m]\in \mathbb{C}^{N}, \quad \text{and} \quad \overline{\vect{w}}_k[m] = \begin{bmatrix} \overline{\vect{w}}_{k1}[m] \\ \vdots \\ \overline{\vect{w}}_{kL}[m] \end{bmatrix}.
\end{align}
Here, $\rho_{\rm max}>0$ is the maximum transmit power allocated per subcarrier at each AP. The normalization by $\omega_k[m]$ ensures compliance with the per-AP power constraint while preserving the direction of the precoding vector, which is computed using global CSI from all APs. Introducing the notation
\begin{align}
    \vect{w}_k[m] = \begin{bmatrix} \vect{w}_{k1}[m] \\ \vdots \\ \vect{w}_{kL}[m] \end{bmatrix},
\end{align}
the transmitted distortion-free signal from AP $l$ at subcarrier $m$ is given as $\overline{\vect{x}}_l[m]= \sum_{k=1}^K\vect{w}_{kl}[m]\overline{s}_k[m]$.

In the distributed operation, the transmitted signal from AP $l$ has the same general structure but differs in the precoding vector computation, as each AP independently determines its precoding vectors based on locally available CSI. A common choice for distributed precoding is local RZF with equal power allocation, given by  
\begin{align}
    \vect{w}_{kl}[m] = \sqrt{\frac{\rho_{\rm max}}{K}} \frac{\overline{\vect{w}}_{kl}[m]}{\left \Vert \overline{\vect{w}}_{kl}[m]\right\Vert}
\end{align}
with
\begin{align}
    \overline{\vect{w}}_{kl}[m] = \left(\sum_{i=1}^K\overline{\vect{h}}_{il}[m]\overline{\vect{h}}_{il}^{\Htran}[m]+\lambda\vect{I}_{N}\right)^{-1}\overline{\vect{h}}_{kl}[m].
\end{align}

The baseband equivalent distortion-free signal transmitted from AP $l$ at the $q$th time-domain sample is given by the $M$-point inverse discrete Fourier transform (DFT) of the sequence $\overline{\vect{x}}_l[m]$, i.e.,
\begin{align}
\vect{x}_l[q] =
\frac{1}{\sqrt{M}}\sum_{m=0}^{M-1}\overline{\vect{x}}_l[m]e^{\imagunit \frac{2\pi q m}{M}}, 
\end{align}
where $q=-(R-1),\ldots,M-1$, and we assume that a CP of length $R-1$ is appended to the time-domain samples. Following the modeling approaches in \cite{Moghaddam2019,Fettweis2005a}, we assume that the hardware imperfections at the APs include residual phase noise and non-frequency-selective IQI. However, since the IQI distortion occurs at the very beginning of the transmit chain and acts as a linear transformation of the real and imaginary parts of the signal, each AP can compensate for this effect locally by adjusting the amplitude and phase of its transmit signal using its knowledge of the IQI parameters. Therefore, without loss of generality, we neglect the impact of IQI at the transmitter side and focus on the effect of the residual phase noise after some phase-tracking scheme such as a phase-locked loop (PLL) \cite{Petrovic2007a,Jacobsson2018}. Then, the input to the power amplifier of AP $l$ UE
\begin{align} \label{eq:s-prime}
\vect{x}_l^{\prime}[q]=e^{\imagunit\psi_l[q]}\vect{x}_l[q].
\end{align}
 Usually, $\psi_l[q]$ is modeled by samples of a first-order recursive stationary process as in \cite{Jacobsson2018} and it is independent of $\vect{x}_l[q]$. We note that the phase noise process has a memory and the correlation between adjacent samples of  $\psi_l[q]$ is $0<\lambda_{\psi_{l}}<1$. It is described as
 \begin{align}
     \psi_l[q] = \lambda_{\psi_l}\psi_l[q-1]+\phi[q]
 \end{align}
 where $\phi[q]\sim \mathcal{N}(0,\sigma^2_{\psi_l})$ with the variance $\sigma^2_{\psi_l}$ being equal to $2\pi\beta_l T_s$, where $\beta_l>0$ is a phase-noise innovation rate parameter and $T_s$ is the sampling period.

 Next, the up-converted signal is passed through the power amplifier. In this paper, we will utilize a third-order quasi-memoryless function where  
both the amplitude and phase of the $n$th entry of the input signal $\vect{x}'_l[q]$, i.e., $x'_{l,n}[q]$ are
distorted as
\begin{align} \label{eq:s-check}
\check{x}_{l,n}[q]=b_{l,1}x'_{l,n}[q]+b_{l,2}\left\vert x'_{l,n}[q] \right \vert^2x'_{l,n}[q],
\end{align} 
where $b_{l,1}\in \mathbb{C}$ and $b_{l,2}\in \mathbb{C}$ are the model parameters specific to the power amplifier of AP $l$. Third-order polynomials with odd-order terms are widely used to model the compression effect of the power amplifiers since the third-order intercept point, which is related
to the third-order term, is a common quality measure for the distortion in RF amplifiers \cite{Ronnow2019}. Quasi-memoryless modeling is meaningful when the bandwidth of the transmit signal is sufficiently low compared to the total bandwidth of the power amplifier \cite{Jacobsson2018, Schenk2008a}. We note that the distortion model in \eqref{eq:s-check} can also be used to represent the residual non-linearities after some predistortion is applied at the APs.

Assuming long-term automatic gain control is used at the AP transmitters \cite{Bjornson2019e,Demir2020}, the power amplifier parameters $b_{l,1}$ and $b_{l,2}$ can be expressed as
\begin{align}
b_{l,1}=\widetilde{b}_{l,1}, \quad \quad b_{l,2}=\frac{\widetilde{b}_{l,2}}{b_{\rm off}\mathbb{E}\left\{\left|x'_{l,n}[q]\right|^2\right\}},
\end{align}
where $\widetilde{b}_{l,1}$ and $\widetilde{b}_{l,2}$ are normalized reference polynomial coefficients adjusted to the signal magnitude between zero and one \cite{3GPP_PA_models}. The parameter $b_{\rm off}\geq 1$ denotes the input backoff applied to the power amplifier at the transmitter of AP $l$ to prevent clipping in the highly compressed region of the amplifier. The overall distorted transmitted signal from AP $l$ is given as
\begin{align}
    \check{\vect{x}}_l[q] = \begin{bmatrix} \check{x}_{l,1}[q] \\ \vdots \\ \check{x}_{l,N}[q]  \end{bmatrix}.
\end{align}
Collecting all the transmitted signals in a vector, we obtain
\begin{align}
    \check{\vect{x}}[q] = \begin{bmatrix} 
\check{\vect{x}}_1[q] \\ \vdots \\ \check{\vect{x}}_L[q] \end{bmatrix} \in \mathbb{C}^{LN}.
\end{align}

\section{Downlink Spectral Efficiency}
In this section, we derive a closed-form lower bound on the downlink capacity for each UE in the presence of hardware impairments. We begin by expressing the received signal at each UE in the time domain and then transform the system model into the frequency domain using the DFT. By applying the Bussgang decomposition, we separate the linear data-dependent component from the distortion introduced by non-linearities. This enables us to compute an effective signal-to-interference-plus-noise ratio (SINR) at each subcarrier and derive a corresponding SE expression. The analysis applies to both centralized and distributed precoding strategies, with the primary difference being the availability of global CSI.

The baseband equivalent of the received signal at UE $k$ is given as
\begin{align}
y_k[q]=\sum_{r=0}^{R-1}\vect{h}_{k}^{\Ttran}[r]\check{\vect{x}}[q-r] + n_k[q],  \label{eq:received-data}
\end{align}
for $q=0,\ldots,M-1$, where $n_k[q]\sim \CN(0,\sigma^2)$ is the additive white complex Gaussian noise. 

Taking the $M$-point DFT of the sequence $y_k[q]$, we obtain
\begin{align} \label{eq:Bussgang-channel-frequency}
\overline{y}_k[m]&=\frac{1}{\sqrt{M}}\sum_{q=0}^{M-1}y_k[q]e^{-\imagunit \frac{2\pi q m}{M}} \nonumber\\
&=\overline{\vect{h}}_k^{\Ttran}[m]\overline{\check{\vect{x}}}[m]+\underbrace{\frac{1}{\sqrt{M}}\sum_{q=0}^{M-1}n_k[q]e^{-\imagunit \frac{2\pi q m}{M}}}_{\triangleq \overline{n}_k[m]\sim \CN(0,\sigma^2)}
\end{align}
for $m=0,\ldots,M-1$. Here, $\overline{\check{\vect{x}}}[m]$ is the $M$-point DFT of the sequence $\check{\vect{x}}[q]$. Now we can analyze the data detection separately at each subcarrier.

We first define $\overline{\vect{s}}[m]=[\overline{s}_1[m] \ \cdots \ \overline{s}_K[m]]^{\Ttran}$. By applying the Bussgang decomposition to $\overline{\check{\vect{x}}}[m]$, where
$\overline{\vect{s}}[m]$ is treated as the input signal \cite{Demir2021}, we obtain
\begin{align} \label{eq:Bussgang-data}
\overline{\check{\vect{x}}}[m]=\underbrace{\mathbb{E}\left\{ \overline{\check{\vect{x}}}[m]\overline{\vect{s}}^{\Htran}[m]\right\}}_{\triangleq\vect{B}[m]}\overline{\vect{s}}[m]+\overline{\bm{\eta}}[m], \quad q=0,\ldots,M-1,
\end{align}
where we use $\mathbb{E}\left\{\overline{\vect{s}}[m]\overline{\vect{s}}^{\Htran}[m]\right\}=\vect{I}_K$.
Here, the expectations are taken with respect to the random data symbols and can be computed using Monte Carlo trials. We refer to $\vect{B}[m]\in \mathbb{C}^{LN\times K}$ as the Bussgang gain matrix. By construction, the distortion term $\overline{\bm{\eta}}[m]\in\mathbb{C}^{LN}$ is uncorrelated with the input signal $\overline{\vect{s}}[m]$. 

The received signal at subcarrier $m$ then can be written as
\begin{align}
\overline{y}_k[m]&= \overline{\vect{h}}_k^{\Ttran}[m]\vect{b}_k[m]\overline{s}_k[m] +
\sum_{\substack{i=1 \\ i \neq k}}^K\overline{\vect{h}}_k^{\Ttran}[m]\vect{b}_i[m]\overline{s}_i[m] \nonumber\\
&\quad +\overline{\vect{h}}_k^{\Ttran}[m]\overline{\bm{\eta}}[m]+\overline{n}_k[m] ,
\end{align}
where $\vect{b}_k[m]$ denotes the $k$-th column of the Bussgang gain matrix $\vect{B}[m]$. By following a similar approach as in the proof of \cite[Thm~6.1]{cell-free-book}, and leveraging the fact that the distortion noise is uncorrelated with the data symbols, we can derive a lower bound on the channel capacity, which represents the maximum data rate under the given channel conditions. This bound corresponds to an SE expression, measured in bits per complex-valued symbol, that is achievable with practical coding and modulation schemes \cite{cell-free-book}. The resulting SE for UE $k$ at subcarrier $m$ is given as
 \begin{align}
 &\mathrm{SE}_{k,m} =  
\log_{2}{(1+\gamma_{k,m})} ,
\end{align}
where $\gamma_{k,m}$ denotes the effective SINR as shown in \eqref{eq:SINR} at the top of the next page,
\begin{figure*}
\begin{align}
\gamma_{k,m} =
\frac
{\left|\mathbb{E}\left\{ \overline{\vect{h}}_k^{\Ttran}[m]
    {\vect{b}}_k[m]\right\} \right|^2}
{\sum_{\substack{i=1 }}^K\mathbb{E}\left\{\left| \overline{\vect{h}}_k^{\Ttran}[m]
    {\vect{b}}_{i}[m] \right|^2\right\}-\left|\mathbb{E}\left\{ \overline{\vect{h}}_k^{\Ttran}[m]
    {\vect{b}}_k[m]\right\} \right|^2
    +\overline{\vect{h}}_k^{\Ttran}[m]\vect{C}_{\overline{\eta}\overline{\eta}}[m]\overline{\vect{h}}_k^*[m]+\sigma^2}. \label{eq:SINR}
\end{align}
\hrulefill
\end{figure*}
where $\vect{C}_{\overline{\eta}\overline{\eta}}[m]=\mathbb{E}\{\overline{\bm{\eta}}[m]\overline{\bm{\eta}}^{\Htran}[m]\}$. This matrix can be computed using Monte Carlo trials. The term ``effective'' SINR implies that the data signal can be encoded and the received signal can be decoded as if communication were taking place over an additive white complex Gaussian noise (AWGN) channel with that SINR. Note that the effective SINR—and consequently, the SE expression—has the same form for both centralized and distributed operations. The key difference lies in the selection of the precoding vectors, which is based on the available CSI: globally in the centralized case, and locally in the distributed case.

\section{Numerical Results}

In this section, we evaluate the downlink SE of cell-free massive MIMO-OFDM systems operating in either centralized or distributed operation, while accounting for hardware impairments including phase noise and third-order power amplifier non-linearities. The simulation setup consists of $L=16$ APs randomly distributed over a $0.5 \times 0.5$\,km$^2$ area. Each AP is equipped with $N=4$ antennas, and there are $K=10$ UEs randomly located within the same region. Path loss in decibels is modeled using the Urban Microcell Street Canyon model as $
-32.4 - 20\log_{10}(f_c) - 31.9\log_{10}(d) + \mathcal{N}(0, 8.2^2)$, where $d$ is the distance between a given AP and UE in meters, $f_c = 7.5$\,GHz denotes the carrier frequency, and the standard deviation of the log-normal shadow fading is 8.2\,dB~\cite[Table 7.4.1-1]{3GPP5G}. A vertical height difference of 10\,m is assumed between the APs and UEs.

The OFDM system uses $M = 256$ subcarriers and $R = 6$ channel taps, with a subcarrier spacing of 15\,kHz. The resulting system bandwidth is $B = 256 \cdot 15\,\text{kHz} = 3.84$\,MHz. The noise variance is computed assuming a noise figure of 7\,dB, yielding $\sigma^2 = -101.16$\,dBm. To model the power delay profile, the Saleh-Valenzuela model~\cite{Saleh1987} is employed with $\Gamma = \gamma = 2$ and five multipath clusters~\cite[Eq.~(7.52)]{bjornson2024introduction}. For each cluster, the azimuth and elevation angles are drawn uniformly at random within a $40^\circ$ neighborhood around the nominal line-of-sight direction. The spatial correlation matrices $\vect{R}_{kl}[r]$ are generated based on the delay profile and uniform linear array (ULA) response vectors.

RZF precoding is applied with the regularization parameter set to $\lambda = 10\sigma^2$, which corresponds to an uplink transmit power of $0.1$\,W in the computation of the uplink-based RZF precoder, following the approach in \cite{cell-free-book}. The maximum downlink transmit power per AP is $\rho_{\rm max} = 2$\,W. Hardware impairments are modeled as follows: third-order power amplifier non-linearities are captured using parameters $\widetilde{b}_{l,1} = 1$, $\widetilde{b}_{l,2} = -1/3$, and a back-off factor $b_{\rm off} = 7$\,dB~\cite{Bjornson2019e}. Phase noise is modeled with parameters $\lambda_{\varphi_l} = 0.99$, $T_s = 1/B$, and $\beta = 10^3$~\cite{Jacobsson2018}.

In the simulation figures, ``cent.'' denotes the SE achieved under centralized operation, while ``dist.'' corresponds to distributed operation. We consider two scenarios: the ``perfect'' case, where hardware impairments are absent, and the ``hardware impaired'' case, which includes the effects of phase noise and/or power amplifier nonlinearities. The cumulative distribution function (CDF) of the SE per subcarrier per UE is plotted. A curve that is shifted further to the right indicates better SE performance, making the corresponding scheme more favorable.

  \begin{figure}[t] 
    \centering
        \includegraphics[width=0.5\textwidth, trim=0.8cm 0.2cm 1cm 0.2cm, clip]{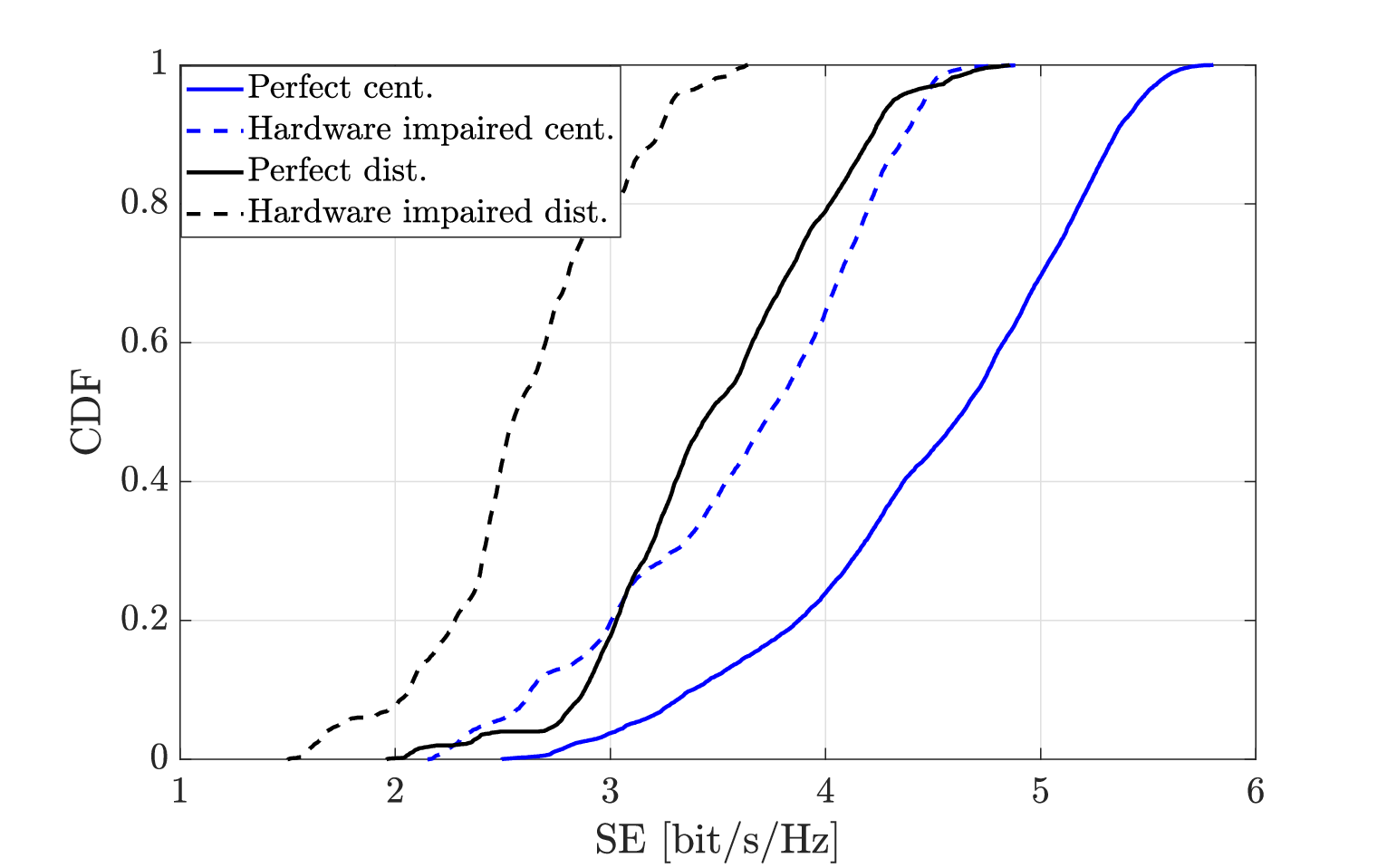} 
         \vspace{-2mm}
        \caption{The CDF of UE SEs when considering perfect hardware and imperfect hardware with centralized and distributed operations. }
        \label{fig:1}
          \vspace{-2mm}
 \end{figure}

  \begin{figure}[t] 
    \centering
        \includegraphics[width=0.5\textwidth, trim=0.8cm 0.2cm 1cm 0.2cm, clip]{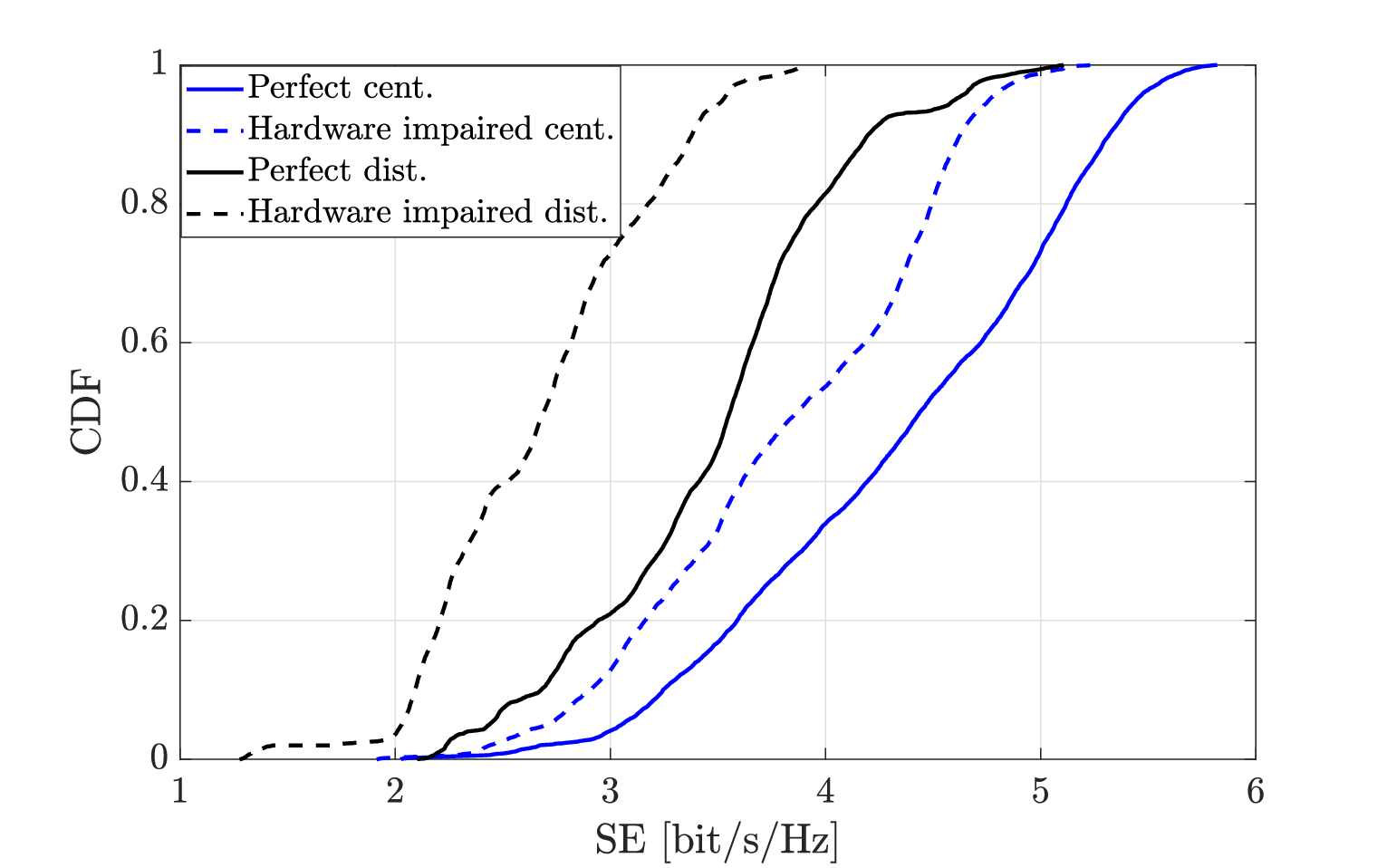} 
         \vspace{-2mm}
        \caption{The CDF of UE SEs when the effect of power amplifier nonlinearities is eliminated and there is only phase noise. }
        \label{fig:2}
          \vspace{-2mm}
 \end{figure}

In Fig.~\ref{fig:1}, we present the CDF of the SE under both phase noise and third-order power amplifier polynomial nonlinearities. The blue curves represent the performance of the centralized operation, while the black curves correspond to the distributed operation. As expected, in both the perfect hardware and hardware-impaired cases, centralized operation achieves higher SE than distributed operation, which aligns with findings in the cell-free massive MIMO literature~\cite{cell-free-book}. This improvement stems from the fact that centralized precoding utilizes CSI from all APs, whereas distributed precoding relies solely on locally available CSI. Another key observation is the noticeable performance degradation in the presence of hardware impairments. This highlights the need for future research on distortion-aware transmit precoding strategies to mitigate the impact of such impairments and close the performance gap.

To evaluate the relative impact of different hardware impairments under the chosen simulation setup, we isolate the effect of phase noise by removing the power amplifier nonlinearities in Fig.~\ref{fig:2}. Compared to Fig.~\ref{fig:1}, a slight performance improvement is observed in the centralized operation, whereas the distributed operation continues to suffer significantly from the presence of phase noise. As a result, when only phase noise is considered, the performance gap between the perfect and hardware-impaired cases becomes smaller for centralized operation compared to distributed operation. We note that these observations reflect the relative impact of impairments for the selected parameter values, which are representative of typical configurations used in prior works. A broader characterization of impairment severity would require evaluating a wider range of parameter settings.

\begin{figure}[t] 
    \centering
        \includegraphics[width=0.5\textwidth, trim=0.8cm 0.2cm 1cm 0.2cm, clip]{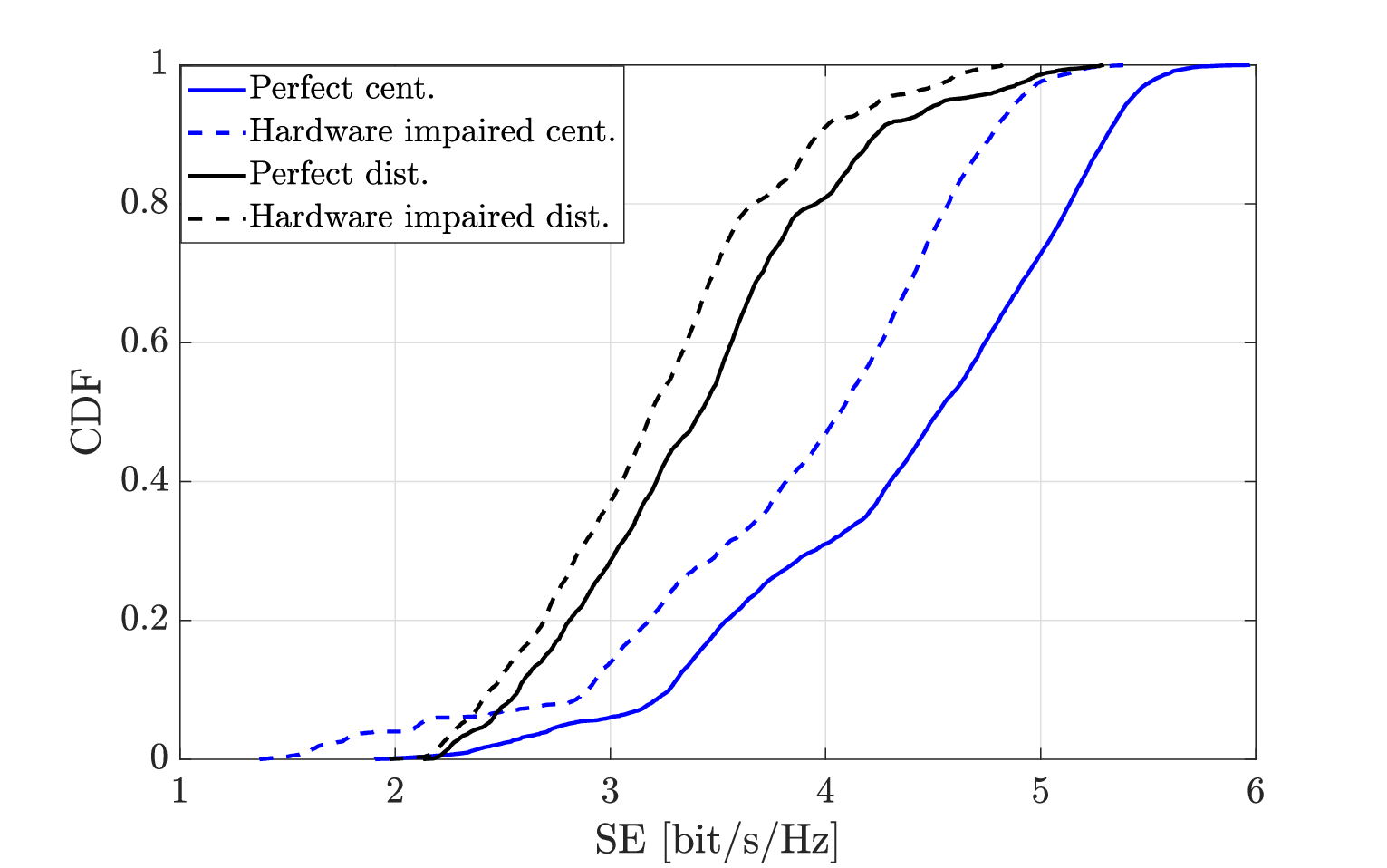} 
         \vspace{-2mm}
        \caption{The CDF of UE SEs when the effect of phase noise is eliminated and there are only power amplifier non-linearities. }
        \label{fig:3}
          \vspace{-2mm}
 \end{figure}

In Fig.~\ref{fig:3}, we retain the impact of third-order power amplifier non-linearities while removing the phase noise, thereby isolating the effect of amplifier distortions. Compared to Fig.~\ref{fig:2}, the hardware-impaired centralized case shows only a negligible change in performance, whereas the distributed operation exhibits a significant improvement due to the elimination of phase noise. These results indicate that for the considered setup, phase noise has a more detrimental impact on system performance than power amplifier non-linearities—particularly in the distributed mode of operation.

\section{Conclusions}

This paper presents a comprehensive downlink performance analysis of cell-free massive MIMO-OFDM systems under practical hardware impairments, focusing on phase noise and third-order power amplifier non-linearities. A detailed system model was developed, covering both centralized and distributed precoding strategies. Using the Bussgang decomposition, we derive a new SE expression.

Our simulation results show that centralized operation consistently outperforms distributed operation in terms of SE, owing to its access to global CSI to suppress the joint interference between the APs. Among the hardware impairments considered, phase noise proves to be the most detrimental, causing more severe performance degradation than power amplifier non-linearities—particularly in distributed configurations. These findings highlight the need for hardware-aware techniques, such as distortion-resilient precoding, to narrow the performance gap relative to ideal hardware. Furthermore, future work could explore the impact of other impairments, such as low-resolution digital-to-analog converters.

\bibliographystyle{IEEEtran}
\bibliography{IEEEabrv,refs}

\end{document}